\documentclass[10pt,twoside]{article}
\usepackage[T1]{fontenc}
\usepackage[applemac]{inputenc}
\usepackage{amssymb,amsmath}
\usepackage[]{graphicx}
\def\volnum{Manuscrit}

\usepackage{aflbcours}
\pdebut{1}

\title{Why Bohr was wrong in his response to EPR}
\titleshort{Why Bohr was wrong in his response to EPR}
\author{Aur\'elien Drezet$^{(1)}$}
\authorshort{A. Drezet}
\address{(1) Univ. Grenoble Alpes, CNRS, Institut N\'{e}el\\
 F-38000 Grenoble, France}
\begin{document}
\maketitle
We assess the analysis made by Bohr in 1935 of the Einstein Podolsky Rosen paradox/theorem.  We explicitly describe Bohr's gedanken experiment involving a double-slit moving diaphragm interacting with two independent particles and show that the analysis provided by Bohr was flawed.  We propose a different protocol correcting Bohr's version that confirms EPR dilemma: Quantum mechanics  is either incomplete or  non-local.    
\vskip 1cm
\begin{abstract}
 
\end{abstract}

\section{Pr\'eambule [hommage \`a Georges Lochak]}
\indent Georges Lochak \'etait un physicien passionn\'e et un \'ecrivain remarquable vulgarisateur magnifique de la physique. Je pense avoir lu tous ses livres mais au moins trois ont marqu\'e ma m\'emoire: il s'agit i) de  \emph{Quanta, Grains et Champs} \cite{Silva} co\'ecrit avec Andrade e Silva qui se focalise sur la pens\'ee de de Broglie et la double solution, ii) sa remarquable biographie de Louis de Broglie intitul\'ee simplement \emph{Louis de Broglie-Un prince de la science} \cite{Lochak1}, livre merveilleux que je conseille \`a toute personne curieuse voulant comprendre l'oeuvre de Broglie (un physicien avec une \^ame d'historien, un chercheur du XIX$^{\textrm{\`eme}}$ si\`ecle exil\'e au XX$^{\textrm{\`eme}}$ \footnote{et cette remarque n'a rien de n\'egative}), et finalement iii) \emph{La g\'eom\'etrisation de la physique} \cite{Lochak2} qui parle de la beaut\'e des lois et symm\'etries en physique. Cependant, Lochak fut avant tout le dernier grand collaborateur de de Broglie (avec Andrade e Silva) et il d\'efendait sa (libre) pens\'ee d\`es qu'il le pouvait contre les critiques qui lui reprochait son retour aux variables cach\'ees (au grand dam des physiciens nombreux en France qui consid\'eraient et consid\`erent encore cela comme r\'etrograde voire m\^eme comme une h\'er\'esie). Je voudrais cependant mentionner un livre qu'il n'a pas \'ecrit mais co\'edit\'e (avec Simon Diner et Daniel Fargue). Intitul\'e \emph{Les incertitudes d'Heisenberg et l'interpr\'etation probabiliste de la m\'ecanique ondulatoire} \cite{Broglie} il s'agit en fait d'un cours in\'edit de Louis de Broglie r\'ealis\'e en 1950-1951. Comme Lochak l'explique dans la pr\'eface, de Broglie ne l'a jamais publi\'e (contrairement \`a son habitude) car:  \begin{quote}
Il l'a d\'esavou\'e, en effet, peu apr\`es l'avoir \'ecrit, car il y exposait, pour la derni\`ere fois sans critique  et d'une mani\`ere particuli\`erement brillante et convaincante, l'interpr\'etation de la m\'ecanique ondulatoire selon les id\'ees de l'\'ecole de Copenhague auxquelles  il adh\'erait encore \`a l'\'epoque mais dont il commen\c{c}a, justement, de douter en relisant son propre texte.\cite{Broglie}
\end{quote} 
Le livre contient entre autre des commentaires de de Broglie et Lochak suite au revirement progr\'essif de la pens\'ee de de Broglie; revirement qui se cristalisa lorsque celui ci re\c{c}u en 1951 un manuscrit de David Bohm \cite{Bohm1952} o\`u l'auteur reprenait en l'\'etendant un peu la vieille th\'eorie de l'onde pilote de 1927 \cite{Broglie1927} (qui devint par la suite la m\'ecanique `Bohmienne'). \\
\indent Pour le pr\'esent article ce qui fondamental c'est que cet ouvrage contient une description magistrale de la th\'eorie de la mesure quantique de l'\'epoque et tout particuli\`erement une analyse des exp\'eriences de pens\'ee propos\'ees par  Heisenberg, Einstein et Bohr dans les ann\'ees 1926-1935. Le c\'el\`ebre d\'ebat Einstein-Bohr qui fit suite aux 5-6$^{\textrm{\`eme}}$ congr\`es Solvay, qui eurent lieu en 1927 et 1930, est analys\'e techniquement en d\'etail par Louis de Broglie. Ce texte est donc pr\'ecieux pour celles et ceux qui ne connaitraient que la perspective de Bohr souvent semi-quantitative et tr\`es difficile \`a suivre. Cependant ce qui nous interessera tout particuli\`erement ici c'est la discussion pr\'ecise du fameux paradoxe d'Einstein Podolsky et Rosen (EPR) publi\'e en 1935 \cite{EPR}. Ce paradoxe fut analys\'e et critiqu\'e par Niels Bohr d\`es 1935 dans le m\^eme journal \cite{Bohr1935} mais sa critique est reput\'ee difficile \`a lire. Tellement en fait que John Bell lui m\^eme affirma qu'il ne comprenait pas ce que Bohr veut dire dans sa r\'eponse \`a	 EPR (\cite{Bell}, chap. 16). Cela motive le pr\'esent travail qui portera sur une analyse plus pouss\'ee qu'a l'usuelle du fameux paradoxe EPR, ou plut\^ot th\'eor\`eme EPR car c'est bien d'un th\'eor\`eme dont il s'agit. C'est la r\'eponse de Bohr qui est ici consid\'er\'ee et critiqu\'ee en s'inspirant de l'analyse faite par de Broglie d'une exp\'erience de pens\'ee propos\'ee par Bohr en 1935 et qui \'etait cens\'ee corriger une erreur d'appreciation et d'analyse des auteurs EPR. Ce travail compl\`ete un peu je pense l'admirable analyse historique faite par de Broglie mais aussi Mara Beller et Arthur Fine \cite{Beller}.    Pour des raisons d'universalit\'e du propos, et en trahissant un peu  ici le souhait  de Lochak de pr\'eserver l'usage de la langue fran\c{c}aise, le reste du pr\'esent article sera r\'edig\'e en anglais.              
\section{EPR: `Einstein attacks quantum theory' (New York Times, May 4 1935)}
\indent It is (un)famously said that Bohr was victorious in the Einstein-Bohr battle; the EPR debate being the culminating point of the fight revealing the grandiose victory of the Copenhagen prophet~\footnote{Einstein famously, in private letters, ironically named  Bohr a `talmudic' or `mystic' philosopher and accused him to behave as a `prophet' preaching complementarity (identified  to a `tranquilizing-pill'). De Broglie   later assimilated Bohr to a `Rembrandt of physics' because of his use of clair-obscur.   } and the last defeat of Einstein `naive' realism. Nothing can be further away of this unfortunate claim repeated again and again in the literature and popular press (especially  with the 2022 Nobel prize of physics attributed to Clauser, Aspect and Zeilinger). To understand this curious issue it is first useful to briefly remind the content of the debate. \\
\indent The EPR paradox/theorem concerns nonlocality and completeness in quantum mechanics. The goal of the EPR article \cite{EPR} was to show that if we assume the principle of `Einstein locality' or EPR-locality (that can precisely defined) quantum mechanics (supposed valid) must be incomplete. More precisely EPR shows that either quantum mechanics is incomplete or quantum  mechanics is nonlocal i.e., it violates Einstein's locality principle.  Moreover, Bohm in his 1952 work \cite{Bohm1952} showed that his own hidden-variables theory able to complete quantum mechanics and based on the pilot-wave theory of de Broglie is explicitly non-local. While this doesn't contradict the EPR results seen as a theorem Bohm approach  was certainly the option `which Einstein would have liked least' (\cite{Bell}, chap.~1). Of course this was just the beginning of the story: In 1964 John Bell, based on EPR work, discovered his famous theorem (\cite{Bell}, chap.~2)  firmly establishing  that quantum mechanics (irrespectively of being complete or incomplete) must be nonlocal.\\
\indent The original EPR article focused on the  position and momentum observables for two entangled particles but it is common  to use instead the example proposed by Bohm in 1951 of two spin-$\frac{1}{2}$ particles 1 and 2 entangled in the singlet state \cite{Bohm1951}. This is not the strategy that we will follow here and we will stick to the original EPR configuration. We remind that the original EPR paper used the two-particle wave function 
\begin{eqnarray}
\Psi(x_1,x_2)=\delta(x_1-x_2-d)\nonumber\\=\iint dk_1dk_2\delta(k_1+k_2) \frac{e^{i(k_1+k_2x_2+k_2d)}}{2\pi}\label{EPR}
\end{eqnarray} in the configuration and momentum spaces showing that we have the strong correlation $x_1=x_2+d$ and $k_1=-k_2$. EPR argued (correctly) that if we measure  the spatial coordinate $x_1$ of the first particle we know counterfactually the spatial coordinate $x_2$ of the second  particle.  All the same we could measure the momentum $k_2$ of the second particle and deduce counterfactually the momentum $k_1$ of the first particle. As a consequence explained EPR we have now precise knowledge  of complementary observables $x_1,x_2,k_1$ and $k_2$ but this is forbidden in the usual complete quantum mechanical interpretation.  Therefore,   assuming locality and the validity of quantum mechanics EPR deduced  quantum mechanics must be incomplete.\\
\indent It is central to understand that for EPR the countefactual deduction is developed from Einstein's locality assumption and not from realism which is not imposed. More precisely,
EPR starts with the Einstein-locality assumption based on natural features of the classical realist World picture concerning  correlations and relativistic causality. The main idea is that a local operation made by Alice on particle 1 at space-time point $x_1$ should not influence what is happening to the second particle recorded by Bob at space-time point $x_2$ if the two events are space-like separated (so that no-signal could propagate between the two points).  As Einstein wrote in 1949 :
 \begin{quote}
 But on one supposition we should, in my opinion, absolutely hold fast: the real factual situation of the system $S_2$ is independent of what is done with the system $S_1$ which is spatially separated from the former.\cite{Einstein1949}, p. 85
\end{quote}  
The crux of EPR is thus to assume that locality or local-causality  is such a natural assumption that we can use it even independently of realistic presuppositions (i.e., without involving hidden variables completing quantum mechanics). This justifies the famous introduction of counterfactual  `elements of reality' by EPR: 
\begin{quote}
If, without in any way disturbing a system, we can predict with certainty (i.e., with probability equal to unity) the value of a physical  quantity, then there exists  an element of physical reality corresponding to this physical quantity.\cite{EPR}
\end{quote}   
\indent Of course EPR understood perfectly well that a counterfactual reasoning is in general forbidden in the usual approach to quantum mechanics for a single particle. This is because   for a single particle one could invoke  Heisenberg's principle to impose a strong form of complementarity and contextuality: It is impossible to record in one single experiment the position $x$ and momentum $k$ because  the operators $\hat{x}$ and $\hat{k}$ don't commute. Therefore, one must choose between one experiment or the other and if one is doing sequential experiments (like measuring $x$, then $k$ and finally again $x$) it is known that dispersion will occur in agreement with Heisenberg's principle.  However, with Einstein-locality EPR found a clean way to somehow circumvent Heisenberg's principle limitations.  Assuming locality we can know counterfactually the complementary variables of the two particles even though we only only  recorded  $x_1$ and $k_2$. Therefore assuming locality we get more information that is usually accepted.  In other words, assuming that  quantum  is local (QM-L) and complete (QM-C) we deduce quantum mechanics must be incomplete (QM-IC)! This is a wonderful logical contradiction that can be formally written:
\begin{eqnarray}
\textrm{QM-L} \Rightarrow   \textrm{QM-IC},&\textrm{ i.e., }&
\textrm{QM-C}  \Rightarrow  \textrm{QM-NL}.
\end{eqnarray} 
where QM-NL (quantum mechanics nonlocal) is the negation of QM-L (i.e., $\lnot$ QM-L).
Importantly,  EPR leads to three physical alternatives:
\begin{eqnarray}
(i) \textrm{QM-L}  & \& &  \textrm{QM-IC}\nonumber\\
(ii) \textrm{QM-NL}  & \& &  \textrm{QM-IC}\nonumber\\
(iii) \textrm{QM-NL}  & \& &  \textrm{QM-C}\label{EPR3}
\end{eqnarray} where (i) was favored by Einstein, (ii) by Bohm (with the pilot wave interpretation), and (iii) correspond to the orthodox interpretation of quantum mechanics assumed by Bohr and followers (even if we will see Bohr didn't clearly get the point). Also it is important to  note that in EPR  incompleteness was obtained through the logical deduction of determinism  meaning that we actually have:  QM-L $\Rightarrow$ QM-D, i.e., quantum mechanics is deterministic \footnote{determinism results from the perfect correlations $x_1=x_2+d$ and $k_1=-k_2$} (assuming locality),  and from that   QM-D $\Rightarrow$ QM-IC.  The beauty and logic of the EPR deduction/theorem is often underappreciated and the fact that counterfactuality and determinism are actually derived and not presupposed by EPR are still nowadays misunderstood.
\section{Bohr's response}     
\indent The detailed answer of Bohr was rather cryptic and lengthy (Einstein in a letter to Schr\"{o}dinger compared Bohr to a talmudic philosopher). Forgetting the cryptic part, and the long reminiscence  concerning  the previous battles he  won against Einstein,  we see that actually Bohr presented an explicit  double-slit example. This involves two initially independent particles impinging with normal incidence on a rigid  but translatable diaphragm with two narrow parallel slits separated by the distance $d$. In his analysis Bohr required the diaphragm to be treated using  quantum mechanics so that we actually have a 3-body quantum system. Bohr didn't give the mathematical details for his device to work but this can be reconstructed aposteriori from his article (see also \cite{Bai,Beller,Broglie,Jammer,Tan,Wooters}). He actually assumed that just after the 3-body interaction the quantum state of the whole entangled system reads:
\begin{eqnarray}
|\Psi\rangle=\iiint dx_1dx_2da\delta(x_1-a)\delta(x_2+d-a)\Phi_1(x_1)\Phi_2(x_2)\frac{e^{iK_0a}}{\sqrt{2\pi}}|x_1,x_2,a\rangle\nonumber\\ \label{Bohr1}
\end{eqnarray}        
 where $a$ is the coordinate of the quantum mechanical diaphragm with initial momentum $K_0$. $\Phi_i(x_i)$ are the transversal wave functions of the two incident particles: These being irrelevant we now write $\Phi_i(x_i)=1$ in the following. Bohr supposes that the `\emph{momentum of this diaphragm is measured accurately before as well as after the passing of the particles'}. This pre and postselection could in principle be easily done using a Compton-like interaction between the massive diaphragm and a long-wavelength photon in order to measure the Doppler effect on the photon frequency (see the Appendix).  Moreover, this implies to consider the  final wave-functions (written in different representations):
 \begin{eqnarray}
\langle x_1,x_2,K|\Psi\rangle=\frac{\delta(x_2+d-x_1)e^{i(K_0-K)x_1}}{2\pi} \label{Bohr2}\\
\langle k_1,k_2,K|\Psi\rangle=\frac{\delta(K_0-K-k_1-k_2)e^{ik_2d}}{2\pi}\label{Bohr3}
\end{eqnarray}  
where $K$ is the postselected  diaphragm momentum\footnote{We have $|K\rangle=\int da\frac{e^{iK a}}{\sqrt{2\pi}}|a\rangle$, $|K_0\rangle=\int da\frac{e^{iK_0 a}}{\sqrt{2\pi}}|a\rangle$ (we use a convention where $\hbar=c=1$).}. From Eqs.~\ref{Bohr2},\ref{Bohr3} we again find the perfect EPR correlation  $x_1=x_2+d$ \footnote{We stress that from Eq.~\ref{Bohr1} we directly get $(\hat{x}_1-\hat{x}_2)|\Psi\rangle= d|\Psi\rangle$ i.e., $d$  is an eigenvalue of $\hat{x}_1-\hat{x}_2$ implying the perfect correlation $x_1=X=x_2+d$. However, $|\Psi\rangle$ is not an eigenstate of $\hat{k}_1+\hat{k}_2$. } and $k_2=K_0-K-k_1$ which reduces to Eq.~\ref{EPR} if $K_0=K$. These conditions actually mean that if we subsequently  measure the coordinates or momentum of the two particles 1 and 2 we obtain the EPR perfect correlations.  \\
\indent At that stage everything would be OK with Bohr's analysis: He provided a mechanical realization of the EPR state starting from a 3-body entangled system and considering a postselection on the momentum $K$ of the diaphragm.  However, in a subsequent step he considered that in order to measure the position of particle 1 it would require 
\begin{quote}
[...] to establish a correlation between its behavior and some instrument rigidly fixed to the support which defines the space frame of reference. Under the experimental conditions described such a measurement will therefore also provide us with the knowledge of the location, otherwise completely unknown, of the diaphragm with respect to this space frame when the particles passed through the slits.
\end{quote} 
In other words, by projecting on a  position state $|X\rangle$ of the mechanical diaphragm we would instead get 
  \begin{eqnarray}
\langle x_1,x_2,X|\Psi\rangle=\delta(x_1-X)\delta(x_2+d-X)\frac{e^{iK_0X}}{\sqrt{2\pi}}\label{Bohr4}\\
\langle k_1,k_2,X|\Psi\rangle=\frac{e^{i(K_0-k_1-k_2)X}}{(2\pi)^{3/2}}e^{ik_2d}\label{Bohr5}
\end{eqnarray}  
where now we have clearly $x_1=X=x_2+d$. Therefore, measuring $X$ (and not $K$) leads to the perfect knowledge of $x_1$ and $x_2$. In turn the price to pay is that we strongly interfered with the postselection procedure leading originally to Eqs.~\ref{Bohr2},\ref{Bohr3}, i.e., to the EPR state (this is not surprising since the operators $\hat{X}$ and $\hat{K}$ do not commute: $[\hat{X},\hat{K}]=i$). From Eq.~\ref{Bohr5} we see that the momentum distribution of particles 1 and 2 is very broad and homogeneous, i.e., $|\langle k_1,k_2,X|\Psi\rangle|^2$ is a constant (compare with Eq.~\ref{Bohr3}). A measure of particle-momentum would not lead to any correlation! Therefore, the new post-selection  on the `pointer' $|X\rangle$ actually precludes the EPR analysis. As Bohr wrote:
\begin{quote}
By allowing an essentially uncontrollable momentum to pass from the first particle into the mentioned support, however, we have by this procedure cut ourselves off from any future possibility of applying the law of conservation of momentum to the system consisting of the diaphragm and the two particles and \textbf{therefore have lost our only basis for an unambiguous application of the idea of momentum in predictions regarding the behavior of the second particle}.
\end{quote}  Bohr's proposal, immobilizing the diaphragm, actually implies that a measure of $x_1$ precludes a measure $k_2$ without disturbance. But this is certainly not what must occur with independent and local measurements of $x_1$ and $k_2$ advocated  by EPR! Therefore, Bohr's example is actually not a fair realization of the EPR scenario. Clearly, Bohr wanted to show that it is not possible to realize the EPR scenario with independent measurements of $x_1$ and $k_2$ but he was actually mistaken as we show below. \\
\indent The rigidity of the diaphragm is a central problem in Bohr analysis. The transfer of momentum is instantaneously transmitted to all parts of the diaphragm even if the slits separation $d$ is supposed very large\footnote{Note that based on this analysis we get $\langle x_1,k_2,K|\Psi\rangle=\frac{e^{i(K_0-K-k_2)x_1}}{2\pi}e^{ik_2d}$, whereas with the Bohr modified state we would instead obtain $\langle x_1,k_2,X|\Psi\rangle=\delta(x_1-X)\frac{e^{i(K_0-k_2)X}}{2\pi}e^{ik_2d}$.  Bob recording conditional probabilities $P(k_2|K)$ (associated with the first protocol) and $P(k_2|X)$ (associated with the second protocol), i.e., averaged on all results of Alice, would  still see a quantitative difference showing the disturbance and invasiveness of Bohr's procedure.}.  But for EPR that was the key point of locality: Avoiding any faster than light communication and influence between \textbf{independent} measurements done on particles 1 and 2.  The device of Bohr cannot properly do the job.  The problem is that by projecting on the diaphragm state $|X\rangle$ (i.e., complementary of the momentum state $|K\rangle$) Bohr didn't play the EPR game: He destroyed the initial EPR state by mixing the postselection and the measurement procedure and therefore precluded the independence of the measurements for $x_1$ and $k_2$. Famously Bohr explained that EPR contains  `\emph{an ambiguity as regards the meaning of the expression without in any way disturbing a system}'. He also wrote that there is `\emph{no question of a mechanical disturbance }' in the experiment protocol but better `\emph{an influence of the very conditions which define the possible types of predictions}'.  All this is not  self-consistent since actually Bohr's proposal for realizing the EPR experiment with a rigid diaphragm clearly contradicts the independence and locality/separation condition required in EPR.     \\
\begin{figure}[h]
\begin{center}
\includegraphics[width=5cm]{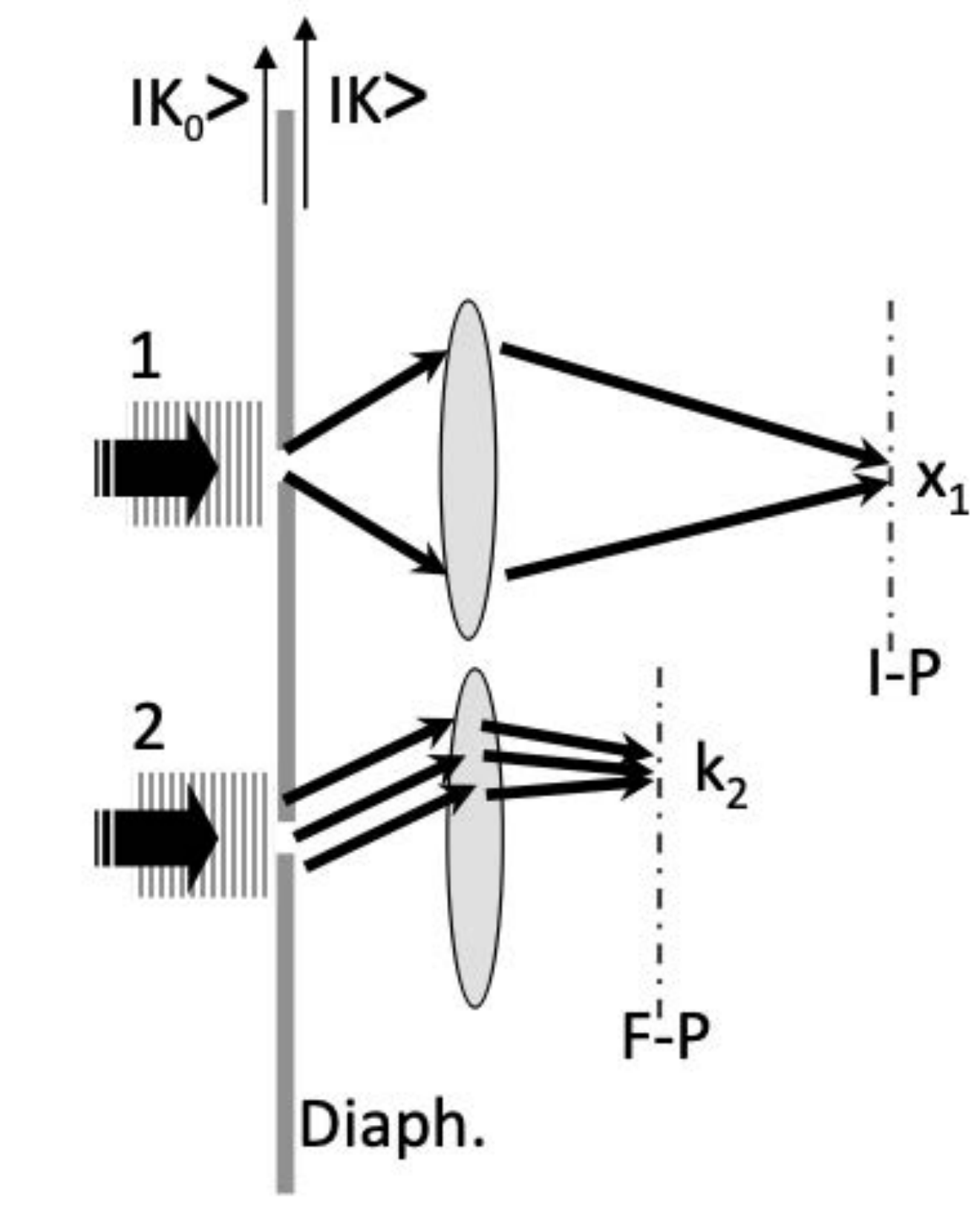} 
\caption{Principle of a gedanken EPR-experiment correcting the proposal of Bohr~\cite{Bohr1935}. Two independent photons 1 and 2 are impinging on a freely moving double-slit diaphragm. The momenta $K_0$ and $K$ of this massive diaphragm are measured before and after the interaction with the two beams using Compton/Doppler effects with additional photons (not shown).  The transverse particle position $x_1$ of photon 1 is imaged using an objective microscope (I-P is the image plane). The  transverse momentum $k_2$ of the second photon is recorded in the back focal plane (Fourier plane : F-P) of a microscope objective (note that in principle only one microscope could be used for both particles).} \label{figure1}
\end{center}
\end{figure}
\indent Moreover, Bohr could have done a correct analysis based on complementarity with his device if he would have preserved the postselection by the diaphragm momentum state $|K\rangle$ leading to the good EPR state. This however requires to not fix the position of the diaphragm (i.e., to not project or postselect  on the state $|X\rangle$).    For example, consider particles 1 and 2 are independent photons transmited by the moving slits:  By using  (see Fig. 1) two optical microscope objectives focused on the diaphragm plane Alice and Bob could decide to image independently the position $x_1$ of the particle 1 (i.e., by recording photons 1 in the image plane of the microscope objective) and the momentum $k_2$ of the particle 2 (i.e., by recording photons 2 in the back-focal plane of the microscope objective). This procedure would not interfere with the post selection leading to the EPR state, i.e., Eqs.~\ref{Bohr2}, \ref{Bohr3}.  The  EPR diagnostic concerning locality  and completeness in quantum mechanics is rigorously valid in this amended gedanken experiment and the deduction of the authors cannot said to be faulty contrarily to Bohr's claims.
\section{Conclusion}
\indent In his later analysis \cite{Bohr1937,Bohr1949} Bohr's emphasized  the indivisibility of quantum phenomena, i.e., `the quantum wholeness' without invoking the details of his 1935 EPR refutation and it has been said that he somehow modified or corrected his first view about complementarity based mostly on local disturbance and the Heisenberg principle (this has been claimed in \cite{Beller}). It is however not very clear that he fully appreciated the real content of the EPR paradox in his writings \cite{Jammer,Beller,Howard}. Clearly, that was not so in 1935. Interestingly, the best argumentation that Bohr could have given was already in the EPR paper where it is written at the end: 
\begin{quote}
One could object to this conclusion on the grounds that our criterion of reality is not sufficiently restrictive. Indeed, one would not arrive at our conclusion if one insisted that two or more physical quantities can be regarded as simultaneaous elements of reality \textrm{only when can be simultaneously measured or predicted}.
\end{quote} This is actually alternative  (iii) of Eq. 3 that  EPR disliked very much since `\emph{No reasonable definition of reality could be expected to permit this}'. It must be observed that Bohr article was not the first response to EPR appearing in print in the physical review. Arthur E. Ruark in a very short paper focused on the previous quotation of EPR and developed the strongly positivist conclusion (iii): 
\begin{quote}
This [EPR] conclusion is directly opposed to the view held by many theoreticians, that a physical property of a given system has reality only when it is actually measured, and that wave mechanics gives a faithful and complete description of all that we can learn from measurements. 
\end{quote}  
Bohr wished a better response: He wanted to convince Einstein the realist of the inevitability of complementarity. Bohr's positivism was a pragmatic one and his goal was to explain why experimental/theoretical considerations tied to  the `\emph{finite interaction between object and measuring agencies}' actually impose his complementarity approach of quantum mechanics that `\emph{fulfill, within its scope, all rational demand on completeness}'. However, the particular  gedanken experiment he provided was flawed and couldn't be used to demonstrate a contradiction. It is also interesting to point out that Bohr didn't explicitly speak  of nonseperability or nonlocality. For Bohr the key feature was indivisibility or impossibility to analyze the quantum system without taking into account the whole experimental setup. Bohr refuted the mere possibility to understand or analyze quantum phenomena within a single causal and spatio-temporal representation that would contradict the spirit of complementarity.   However, this issue was not in conflict with the EPR result since alternative (iii) -considered by Einstein as non plausible- was clearly discussed in \cite{EPR}. \\
\indent  Retrospectively can we say that Einstein won the debate against Bohr? Alas, this is not so simple because actually the alternative (i) that Einstein  favored was later shown to be untenable by Bell in 1964:   Quantum mechanics must be nonlocal irrespectively of being complete or incomplete. Still, it must be said that locality or more precisely local-causality is a compound principle merging   locality and causality, parameter setting and outcome independence. There are several possible loopholes that can be used in order to built up hidden-variables theories agreeing with the results of Bell's theorem  and give sense to the spooky nonlocality.     
\section{Appendix}
\indent In his book \emph{Les incertitudes d'Heisenberg et l'interpr\'etation probabiliste de la m\'ecanique ondulatoire} \cite{Broglie} de Broglie discussed the measurement of the velocity of an electron using the Doppler effect. The idea is to consider a photon-electron collision. If $\omega$, $\omega'$ (respectively $v$, $v'$ ) denote the photon energy (respectively electron velocity) before and after the collision we have 
\begin{eqnarray}
\omega+\frac{1}{2}mv^2=\omega'+\frac{1}{2}mv'^2\nonumber\\
mv-\omega=mv'+\omega'.
\end{eqnarray} 
From this we deduce
\begin{eqnarray}
\omega-\omega'=\frac{1}{2m}[(\omega+\omega')^2-2mv(\omega+\omega')]
\simeq 2\frac{\omega^2}{m}-2v\omega
\end{eqnarray} containing a Compton and Doppler recoil. In the limit where the electron mass is very large we obtain $\omega-\omega'\simeq -2mv\omega$ which is Doppler formula. The result is left unchanged if we replace the electron by the diaphragm with very large mass $M$. Therefore, measuring the photon frequency shift while illuminating the quantum diaphragm we can deduce its  momentum $K_0$ and $K'$ before and after the interaction with the two particles 1 and 2.
\vskip 30pt
\begin{eref}
\bibitem{Bai}
T. Bai, J. Stachel, Bohr's diaphragms in Quantum structural studies, eds. R. Kastner and G. Jaroszkiewicz, pp. 23-52 (World Scientific, 2017). 
\bibitem{Guido}
G. Bacciagaluppi, Did Bohr understand EPR? in One Hundred Years of the Bohr Atom, eds. F. Aaserud and H. Kragh, Scientia Danica. Series M, Mathematica et physica, Vol. 1, pp. 377-396 (Royal Danish Academy of Sciences and Letters, Copenhagen, 2005). 
\bibitem{Bell}
J.~S. Bell, Speakable and unspeakable in quantum mechanics, second edition, Cambridge University Press, Cambridge (2004).
\bibitem{Beller}
M. Beller, Quantum dialogue (University of Chicago Press, Chicago, 1999); M. Beller and A. Fine Bohr's response to EPR in Niels Bohr and contemporary philosophy, eds. J. Faye and H.J. Folse, pp. 1-31 (Kluwer, Dordrecht, 1994).
\bibitem{Bohr1928}
N. Bohr, The quantum postulate and the recent developement of atomic theory, Supplement to Nature, 580-590  (april 14, 1928).
\bibitem{Bohr1935}
N. Bohr, Can quantum-mechanical description of physical reality be considered complete?, Phys. Rev. \textbf{48}, 696-702 (1935); Quantum mechanics and physical reality, Nature (London) \textbf{136}, 65 (1935).
\bibitem{Bohr1937}
N. Bohr, Causality and complementarity, Philosophy of Science \textbf{4}, 289-298 (1937).
\bibitem{Bohr1949}
N. Bohr, Discussion with Einstein on epistemological problems in Atomic Physics, in Albert Einstein Philosopher scientist, edited by P.A. Schilpp, Open Court; 3rd edition (1998).
\bibitem{Bohm1951}
D. Bohm, Quantum Theory (Prentice-Hall, New York, 1951).
\bibitem{Bohm1952}
D. Bohm, A suggested interpretation of the quantum theory in terms of hidden variables-I and II, Phys. Rev. \textbf{85}, 166-179; 180-193 (1952).
\bibitem{Broglie1927}
L. de Broglie, La m\'ecanique ondulatoire et la structure atomique de la mati\`ere et du rayonnement, J. Phys. Radium \textbf{8}, 225-241 (1927). 
\bibitem{Broglie}
L. de Broglie, Les incertitudes d'Heisenberg et l'interpr\'etation probabiliste de la m\'ecanique ondulatoire, eds. S. Diner, D. Fargue, G. Lochak (Gauthier-Villars, Paris, 1982).
\bibitem{EPR}
A. Einstein, B. Podolsky, N. Rosen, Can quantum-mechanical description of physical reality be considered complete?, Phys. Rev. \textbf{47}, 777-780 (1935).
\bibitem{Einstein1949}
A. Einstein, Autobiographical notes, in Albert Einstein Philosopher scientist, edited by P.A. Schilpp, Open Court; 3rd edition (1998).
\bibitem{Howard}
Don Howard, Nicht sein kann was nicht sein darf, or the prehistory of EPR, 1909-1935: Einstein's early worries about the quantum mechanics of composite systems, 
in Sixty-two years of Uncertainty,  ed. A.I. Miller, pp. 61-111 (Plenum Press, New York, 1990); Revisiting the Einstein-Bohr dialogue, Iyyun: The Jerusalem philosophical  quaterly \textbf{56}, 57-90 (2007).  
\bibitem{Jammer}
M. Jammer, The philosophy of quantum mechanics (Wiley, New York, 1974). 
\bibitem{Lochak1}
G. Lochak, Louis de Broglie - un prince de la science  (Flammarion, Paris, 1999).
\bibitem{Lochak2}
G. Lochak, La g\'eom\'etrisation de la physique (Flammarion, Paris, 2013).
\bibitem{Paty}
M. Paty, The nature of Einstein's objections to the copenhagen interpretation of quantum mechanics, Found. Phys. \textbf{25}, 183-204 (1995).
\bibitem{Ruark}
A. E. Ruark, Is the quantum-mechanical description of physical reality complete? Phys. Rev. \textbf{48} (Letters to the editor) 466-467 (1935).
\bibitem{Silva}
J.~L. Andrade e Silva, G. Lochak,  Quanta, Grains et Champs (Hachette, Paris, 1969).
\bibitem{Tan}
S.~M. Tan and D. F. Walls, Loss of coherence in interferometry, Phys. Rev. A. \textbf{47}, 4663-4676 (1993).
\bibitem{Wooters}
W.~K. Wooters, W.~H Zurek, Complementarity in the double-slit experiment: Quantum nonseparability and a quantitative statement of Bohr's principle, Phys. Rev. D. \textbf{19}, 473-484 (1979). (1993).  
\end{eref}
\end{document}